\documentstyle[aps,prl,preprint]{revtex}
\begin{document}
\draft
\title{Crossover from $O(3)$ to $O(4)$ behavior in weakly frustrated
antiferromagnets.}
\author{Andrey V. Chubukov${}^{1,2}$ and Oleg A. Starykh${}^{3,}$~\cite{oas}}
\address{
${}^{1}$Department of Physics, University of Wisconsin, Madison, WI 53706\\
${}^{2}$P.L. Kapitza Institute for 
Physical Problems, Moscow, Russia\\
${}^{3}$ University of California at Davis, Davis, CA 95616}
\date{today}
\maketitle
\begin{abstract}
We consider an anisotropic version of the $CP^{1}$ model which describes
frustrated quantum antiferromagnets with incommensurate spin correlations.
 We extend the two-component spinon field, describing lattice spins,
to the $M$-component complex vector, and show, in the $1/M$ expansion,
that for arbitrary small incommensurability
longitudinal and transverse stiffnesses tend to the same value
as the system approaches the quantum critical point. 
For physical spins ($M=2$), this yields $O(4)$ critical behavior.
However, if the spin structure is commensurate, 
the longitudinal stiffness is identically zero. In this case, the critical
behavior is the same as in  $O(3)$ sigma model. 
We show how the critical exponents interpolate between $O(3)$ and $O(4)$
values near the transition. We also show that the competition between these two
fixed points leads to a confinement-deconfinement transition at
a finite temperature.
\end{abstract}

In this communication, we address a specific issue concerning the
nature of zero-temperature quantum 
phase transitions in quantum antiferromagnets:
whether there
is a qualitative change in the critical behavior when  one 
adds a frustration to Heisenberg antiferromagnet and makes spin configuration
non-collinear.  We argue that the critical behavior of collinear and
non-collinear antiferromagnets is qualitatively different, even when local spin
configuration differs from a Neel state by an arbitrary small amount. 
As a quantitative measure of this difference, we consider the ratio of the
fully renormalized 
transverse and longitudinal stiffnesses, $\rho_{\perp}/\rho_{\parallel}$. In
a collinear antiferromagnet, the 
longitudinal spin stiffness is identically zero because the
Neel ordering is described by just one vector of antiferromagnetism, ${\vec
n}$, and a rotation of ${\vec n}$  around its
equilibrium direction is {\it not} a symmetry transformation. 
 On the other hand,
in the ordered non-collinear state, one needs two unit vectors to
describe the equilibrium spin configuration. In this case, a rotation
of one vector around another is a legitimate 
symmetry transformation which is broken in the ordered state. As a result,
there are  two finite stiffnesses: 
one for two equivalent transverse spin-wave modes,  and another for the 
longitudinal spin-wave mode. The action for non-collinear antiferromagnets 
can be written either in terms of the $SO(3)$ rotation matrix, or in terms
of two-component complex spinon field. Azaria et al. performed one-loop RG
 studies of the $SO(3)$ action~\cite{az} and found that as the system 
approaches the critical point, the ratio of the fully renormalized stiffnesses
tens to unity. 
 Sachdev,  Senthil and one
of us used the spinon description and
 have studied the critical behavior of frustrated antiferromagnets under
the assumption that the bare transverse and longitudinal 
stiffnesses, $\rho^{0}_{\perp}$ and $\rho^{0}_{\parallel}$, 
are close to each other~\cite{CSS}. They also found that
as the system approaches the critical point, the ratio of the fully
renormalized stiffnesses tends to unity.
At criticality, $\rho_{\perp} = \rho_{\parallel}$, and the symmetry of the
underlying action is enlarged from $SU(2) \times U(1)$ 
to $O(4)$ (see below). In this paper, we extend the spinon approach to an 
arbitrary ratio of the bare stiffnesses.  We will show
 that the $O(4)$ critical behavior holds for arbitrary
$\rho^{0}_{\parallel} / \rho^{0}_{\perp} \leq 1$. 
We cannot say at the 
moment about how far the $O(4)$ behavior extends to a region where 
$\rho^{0}_{\parallel} \gg \rho^{0}_{\perp}$.

Our point of departure is the partition function for a frustrated
antiferromagnet written in terms of spinon fields, 
${\cal Z}=\int D{z^*} D{z}  \exp {[-S]}$, where
 \begin{equation}
S = 2{\rho}^{0}_{\perp} \int d{\tau} d^2{\bf r} \left[
|\partial_{\mu}z|^2 - \frac{\gamma}{4} (z^* \partial_{\mu}z -
z \partial_{\mu}z^*) ^2\right].
\label{zzz}
\end{equation}
Here $z$ is a two-component complex spinor field 
subject to constraint
$z_{\alpha}^* z_{\alpha} =1$, ${\mu}={\tau}, x, y$, and $\gamma =
(\rho^0_{\parallel} - \rho^0_{\perp})/\rho^0_{\perp}$.
 For simplicity we choose
units where $\hbar =1$ and set both spin-wave velocities  to unity.

This effective action 
can be explicitly derived from semiclassical 
microscopic considerations~\cite{dombre-read,angel}
 and the general macroscopic approach of
Ref~\cite{andreev}. Note however that $z$ quanta are 
{\it not}  Schwinger
bosons. The relation between $z$ and the underlying spins is more  complex 
and involves 
incommensurate ordering momentum~\cite{angel,CSS}. The action in (\ref{zzz})
is invariant under global $SU(2)$ spin
rotation, and is also invariant under certain type of lattice transformations.
As shown in ~\cite{CSS}, this lattice symmetry is in essence identical to a
lattice $U(1)$ symmetry. We will thus refer to 
the total global symmetry of the  effective action in (\ref{zzz})
as  $SU(2) \times U(1)$. It is essential however, that the $U(1)$ gauge
symmetry $ z({\vec r}, \tau) \rightarrow z({\vec r}, \tau)e^{i \phi ({\vec r},
\tau)}$ is broken provided $\rho_{\parallel} \neq 0$, i.e., 
$\gamma > -1$.  It is only present at $\gamma = -1$ in which case
the action in (\ref{zzz}) describes collinear antiferromagnets. In this latter
case, the description in terms of $z$ quanta is equivalent to 
the description in
terms of Schwinger bosons: the same action as in (\ref{zzz})
 is  obtained if one 
introduces the $U(1)$ gauge invariant Schwinger boson decomposition 
$n^a = b^{\dagger}_{\alpha} \sigma^a_{\alpha \beta} b_{\beta}$   
into the
partition function of the $O(3)$ ${\vec n}-$field model~\cite{CSS,AO}. 
$\gamma \ll 1$ limit of the action (\ref{zzz}) 
was studied in detail in Ref.~\cite{CSS}.  
Here we focus on a region near $\gamma = -1$. 

To perform $1/M$ expansion, we need to generalize the action to large $M$.
We generalize the doublet $z$ to the M-component complex vector,
rescale the $z$ field to $z\rightarrow z/\sqrt{M}$ 
(such that $z_{\alpha}^* z_{\alpha} =1$, $\alpha = 1,2 ...M$), and 
 introduce the coupling constant
$g=M/2{\rho}^0_{\perp}$. We further introduce 
the Hubbard-Stratonovich vector gauge field $A_{\mu}$ to decouple
the quartic term, and introduce a constraint into the action using the integral
representation of the $\delta-$function. We then obtain
\begin{eqnarray}
&&S=\frac{1}{g}\int d^2{\bf r} \int_0^{1/T} d{\tau} {\cal L},
\\
\label{D1}
&&{\cal L}=[(\frac{1}{1 +r}|(\partial_{\mu} -iA_{\mu})z|^2 + 
\frac{r}{1+r} |\partial_{\mu}z|^2 + 
 i\lambda(|z|^2-M)]
\nonumber
\end{eqnarray}
where we introduced $r = -(1 + \gamma)/\gamma$.

The most straightforward way to compute the ratio of the fully renormalized
stiffnesses, which we will follow, is to perform calculations in the ordered
state at $T=0$. This state is realized for $g$ smaller than the 
critical coupling $g_c$.
Assume that the first component of $z$ is condensed. We then
write $z = ({\bar \sigma},\pi_{\alpha})$, and represent ${\bar \sigma}$ as  
a sum of the condensed part, $\sqrt{M}\sigma_0$, and fluctuation 
$\sigma$ around it, 
${\bar \sigma} = \sqrt{M} \sigma_0 + \sigma$. It is also convenient to
introduce pairs of real variables instead of complex variables $\sigma$
and $\pi_{\alpha}$: $\sigma = \chi + i \eta,~~\pi_{\alpha} =
\phi_{2\alpha -1} +i \phi_{2 \alpha}$, and 
rescale gauge field as ${A_{\mu}} \rightarrow (1 +r) {A_{\mu}}$. 
Substituting these expressions into the
action we find
\begin{eqnarray}
&&{\cal L}= \biggl[(\partial_{\mu}\chi)^2 + (\partial_{\mu}\eta)^2 + 
\sum_{\alpha =1}^{2M-2}
(\partial_{\mu}\phi_{\alpha})^2 - 2 \sqrt{M} \sigma_0 \eta 
 \partial_{\mu}A_{\mu}\nonumber \\
&& - 2 A_{\mu} (\chi 
\partial_{\mu}\eta - \eta \partial_{\mu}\chi + \phi_1 \partial_{\mu}\phi_2 -
\phi_2  \partial_{\mu}\phi_1 + ...)  + \nonumber \\
&& A^2_{\mu} M (1+r) 
 + i \lambda (2 \sqrt{M}
\sigma_0 \chi + \chi^2 + \eta^2 + \sum_{\alpha =1}^{2M-2}
\phi^2_{\alpha})\biggr].
\label{D2}
\end{eqnarray}

This decomposition of the $z$ field implies that 
the variables $\phi_{\alpha}$
describe $2M-2$ transverse fluctuations, $\eta$ is a variable for a 
longitudinal mode, and $\chi$ describes fluctuations in the direction
of the condensate.

Our first goal is to integrate out fluctuations of  
$\chi, ~\eta$, and
$\phi_{\alpha}$ and to obtain the effective action for collective 
variables ${\bf A}$ and $\lambda$.
The integration over the longitudinal
and transverse fluctuations yields contributions to the effective action which
are linear in $M$, whereas the $\chi$ field contributes only a
 subleading, $0(1)$, term which can be safely neglected in the leading
 order calculations in $1/M$. 
Performing the Gaussian integration over fluctuating fields and using the
constraint equation at $M = \infty$, we obtain 
the effective action for the gauge fluctuations in the form
$S_A = (M/2) \int d^3 {\bf q} ~\Pi_{\mu \nu} (q)~A_{\mu}(q) A_{\nu}(-q)$, where
\begin{eqnarray}
&&\Pi_{\mu \nu} (q) =  2\int \frac{d^3 k}{(2\pi)^3} 
\frac{1}{k^2} - \int \frac{d^3 k}{(2\pi)^3}
\frac{(2k_{\mu} + q_{\mu}) (2k_{\nu} + q_{\nu})}{k^2 (k+q)^2}
\nonumber \\
&&+ ~2\delta_{\mu \nu} \frac{2(r + \sigma_0^2)}{g} - ~\frac{2\sigma_0^2}{g}
\frac{q_{\mu}q_{\nu}}{q^2}~.
\label{eff}
\end{eqnarray}
The first two terms in (\ref{eff}) are the components of the
polarization operator of the $O(3)$ sigma model, $\Pi^{\phi}_{\mu\nu}(q)$, 
which should be massless due to the gauge invariance of the latter.
Using the Pauli-Willars regularization, we obtain
$\Pi^{\phi}_{\mu \nu}(q)=(\delta_{\mu \nu} - \frac{q_{\mu} q_{\nu}}{q^2})
\frac{q}{16}$.
Collecting all terms in (\ref{eff}) and inverting the result
we find for the gauge field propagator
\begin{eqnarray}
D_{\mu \nu}(q)=&&\frac{1}{M}\Biggl((\delta_{\mu\nu}-\frac{q_{\mu}q_{\nu}}{q^2}
)\left[\frac{q}{16} + \frac{2(r + \sigma_0^2)}{g}\right]^{-1} 
\nonumber \\
+ &&~\frac{q_{\mu} q_{\nu}}{q^2} \frac{g}{2r} \Biggr)
\label{gauge}
\end{eqnarray}
Notice that the longitudinal part of the propagator appears only due to
 incommensurability \cite{ital2}. This is a direct
consequence of the fact that incommensurability breaks the
 gauge symmetry of the action (\ref{D1}).

The  propagator of the constraint field, $\Pi^{-1}(q)$,  
can be calculated in a similar way.
 We found that, to leading order in $1/M$, $\Pi (q)$ 
is independent on $r$ and has the same form as in 
~\cite{CSS}:
$\Pi^{-1}(q)= {8 q^2}/(q + 16 \sigma_0^2/g)$.

 We proceed now with the calculations of the stiffnesses at $M= \infty$.
 The two stiffnesses can be 
extracted from the long-distance behavior of the propagators of the transverse
and longitudinal fields: $G^{-1}_{\phi_{\alpha}} (q) = \rho_{\perp} q^2,~ 
G^{-1}_{\eta} (q) = \rho_{\parallel} q^2$.
The computation of these propagators is straightforward.
Transverse fields $\phi_{\alpha}$
 do not directly couple to the condensate, and
all corrections to the free-particle propagator 
have relative $1/M$ smallness. Hence full $M=\infty$ propagator coincides
with the bare one 
$ G_{\phi_{\alpha}}(q)= g/(2q^2)$, i.e., $\rho_{\perp} = 2/g$.
The longitudinal ($\eta$-field) propagator is however 
different from $G_{\phi_{\alpha}}$ already at $M = \infty$ because
of the coupling term 
$\propto \sqrt{M}\sigma_0 \eta \partial_{\mu} A_{\mu}$ in (\ref{D2}).
This term leads to the finite $M=\infty$ correction,
\begin{equation}
G_{\eta}(q)=\frac{g}{2q^2} + M\sigma_0^2~\frac{q_{\mu} q_{\nu}}{q^4}
D_{\mu \nu}(q)=\frac{g}{2q^2}~\frac{r + \sigma_0^2}{r},
\label{longit}
\end{equation}
so that $\rho_{\parallel}=\frac{2r}{g(r + \sigma_0^2)}$. 
We emphasize that Eq.(\ref{longit}) is an exact
$M=\infty$ result, not an expansion around the 
free-particle expression.        

For the ratio of the fully renormalized stiffnesses we then obtain
\begin{equation}
{\bar \gamma} = \frac{\rho_{\parallel} - \rho_{\perp}}{\rho_{\perp}} = 
- \frac{\sigma_0^2}{r + \sigma^2_0} \equiv \gamma ~\frac{\sigma_0^2}{ 1 +
 \gamma - \gamma \sigma_0^2}
\label{ratio}
\end{equation}
This is the key result of the $M=\infty$ consideration.
 We see that as long as the longitudinal stiffness is finite (i.e., 
$\gamma > -1$),
 the ratio of the fully renormalized stiffnesses approaches $0$ as the
system moves to the critical point, $\sigma_0 \rightarrow 0$.
 At criticality, ${\bar \gamma} =0$, and
 the renormalized action (\ref{zzz})  reduces to that for the $O(2M)$ sigma
model. At the same time, if $\rho^0_{\parallel}
=0$ (i.e., $\gamma =-1$), then ${\bar \gamma} = \gamma = -1$,
and the renormalized action retains the symmetry of the isotropic 
$CP^{M-1}$
model. Alternatively stated, for $\rho_{\parallel} =0$, 
which is
the case for a collinear antiferromagnet, 
 the system is in the basin of
attraction of the $CP^{M-1}$ fixed point. However, an arbitrary small 
amount of non-collinearity drives the system to the $O(2M)$ fixed point.

Our next goal is to compute 
 $1/M$ corrections to the ratio of stiffnesses. These
corrections come from self-energy diagrams which involve exchange by 
fluctuations of both the constraint and the gauge field.  
The computational steps are rather involved, but conceptually
are similar to those discussed in 
Refs.~\cite{CSY,CSS}.
For $r \gg \sigma^2_0$ we obtained 
\begin{eqnarray}
&& G^{-1}_{\phi} = \frac{2q^2}{g}~\left(1 - \frac{4}{3 \pi^2
M}~L\right) \\ 
&&G^{-1}_{\eta} = \frac{2q^2}{g}~\left(1 - \frac{4}{3 \pi^2 M}~L\right)
\left(1 - \frac{\sigma^2_0}{r} (1 + \frac{28}{3 \pi^2 M}~L) \right)
\nonumber
\end{eqnarray}
where $L = log (g_c - g)/g_c$, 
and $\sigma_0$ is related to $g_c -g$ as in the $O(2M)$ model ~\cite{CSY},
$\sigma^2_0 = (1 -g/g_c)^{1 -4/(\pi^2 M)}$.
This is merely a consequence
of the fact that the propagator of the constraint field does not depend on $r$.
For the ratio of the stiffnesses we then obtain 
${\bar \gamma} = (\gamma/(1 + \gamma))~
\left(1-g/g_c\right)^{1 + 16/(3 \pi^2 M)}$.
For $\gamma \ll 1$, this reproduces the result of Ref.~\cite{CSS}. 
We see that the $1/M$ corrections only speed up the flow to the $O(2M)$ fixed
point~\cite{comment}.
  For $r \leq \sigma^2_0$, the expression for ${\bar \gamma}$ is rather
involved and we refrain from presenting it. 
                               
For completeness, we also computed critical exponents in the $1/M$
expansion. For $\gamma = -1$ (i.e., $r=0$) spinon correlation function and
order parameter possesses
$\eta = -\frac{20}{\pi^2 M}, ~~~\nu = 1 + \frac{16}{\pi^2 M},~~~2\beta = 1 -
\frac{4}{\pi^2 M}$.
 The actual spin susceptibility is a convolution of two spinon fields,
and it has different critical exponents ${\bar \eta}, {\bar \nu}$, etc.
 We computed the spin susceptibility  to order $1/M$ and found
${\bar \eta}  = 1 -\frac{32}{\pi^2 M}, ~~~{\bar \nu} = {\nu},~~
{\bar \beta} = 1 + O\left(\frac{1}{M^2}\right).$
The result for ${\bar \eta}$ has been reported by us
previously~\cite{AO}.

For $\gamma > -1$ (i.e., $r >0$), the gauge field acquires a mass,
 and the self-energy terms associated with 
the exchange of gauge field
fluctuations are no longer singular. We have checked that the
spinon fields now possess $O(2M)$ exponents~\cite{CSY}:
$\eta = \frac{4}{3\pi^2 M}, ~~~\nu = 1 - \frac{16}{3\pi^2 M},~~~2\beta = 1 -
\frac{4}{\pi^2 M}$.
 For spin-spin correlation function
and magnetization (at arbitrary $\gamma$) 
we reproduced the results of Ref~\cite{CSS}:
${\bar \beta} = 1 + O\left(\frac{1}{M^2}\right), ~~{\bar \nu} = 1 -
\frac{16}{3 \pi^2 M}, ~~{\bar \eta} = 1 + \frac{32}{3 \pi^2 M}$.

 So far we were discussing the zero-temperature critical properties of
incommensurate antiferromagnets.  As the exponent ${\bar \eta}$ is finite
for both $\gamma=-1$ and $\gamma > -1$,
the staggered static spin susceptibility, $\chi (q)$, at criticality  
possesses a branch-cut singularity independent on whether the system is in the
basin of attraction of $O(3)$ or $O(4)$  fixed points. ~From this 
perspective, the difference between $O(3)$ and $O(4)$ critical
behavior at $T=0$ is only quantitative but not qualitative one.
 In general, however, the descriptions of
the system in terms of ${\vec n}-$field and in terms of spinons are
fundamentally different: in the first case the excitations necessary possess
integer spin while in the latter one can have 
 excitations with either integer or half-integer spin
depending on whether spinons are confined or not.
We now show that the presence of two fixed points at $T=0$ gives rise to
a confinement-deconfinement transition within the {\it{disordered}} region.
Consider for definiteness 
the renormalized - classical (RC) regime, $g < g_c$~\cite{CHN}.     
Here the ground state is ordered, 
but an arbitrary
small temperature leads to
the restoration of the spin rotational symmetry which implies
that one can no longer distinguish between 
transverse and longitudinal fluctuations. Consider first collinear
antiferromagnet, $\gamma =-1$. Then the critical behavior is governed by
the $O(3)$ exponents. 
As the  isotropic $CP^1$ model is isomorphic to
the $O(3)$ sigma-model, one should obtain the same result within the 
${\vec n}$-field and the spinon
description. We recently demonstrated~\cite{AO} that this is indeed the case.
Despite the fact that 
effective action (\ref{zzz}) yields 
a branch cut behavior of $\chi (q)$ at the mean-field ($M \rightarrow \infty$) 
level,
the gauge field fluctuations, which appear at the $1/M$ level,
 confine spinons into pairs with integer spin.
 These bound states of spinons yield poles in $\chi (q)$, and 
the long-distance behavior of spin correlators totally consistent with the 
${\vec n}$-field  description.  The
 branch-cut to pole transformation 
is a direct consequence of 
the gauge-invariance of the action (\ref{zzz}) at $\gamma =-1$. 
Unbroken gauge invariance  leads to a
gapless gauge field fluctuations, which give rise to the unbounded
long-range confining potential between spinons.

Let us consider now what happens at $\gamma > -1$.
We have shown in ~\cite{AO} that the
 static staggered susceptibility $\chi (q)$
is  proportional to the Green's function
$\Psi(q, x=0)$ of the effective inhomogeneous Schrodinger equation 
\begin{equation}
(-\frac{d^2}{dx^2} + V(x) + \delta^2)\Psi(x)=\delta(x),
\label{schrod}
\end{equation}
where $\delta^2=\frac{q^2}{4} + m_0^2$, and $m_0$ is the gap in
the spinon spectrum. At $M = \infty$, the confining potential
$V(x)$ vanishes, and $\chi (q)$ has a 
branch-cut singularity  at $\delta =0$. At finite $M$, 
 $V(x)$ 
is given by the regularized Fourier transform
of the transverse part of the gauge field propagator~\cite{AO}
\begin{equation}
V(x)=\frac{m_0 T}{M} \int_{-\infty}^{\infty} \frac{d k}{2\pi} D_{trans}(k)
(1 - e^{-ikx}).
\label{potential}
\end{equation}

In the RC regime, 
we found that in the small momentum limit, $D_{trans} (k)$ is
\begin{equation}
D_{trans}(k)=\frac{1}{M}\left[\frac{k^2 T}{12\pi m_0^2} + 
\frac{2r}{g}\right]^{-1}.
\label{trans}
\end{equation}
 The appearance of the mass term $2r/g$ in $D_{trans}(k)$ makes the
would-be-confining potential between spinons short-ranged:
 $V(x) = (6\pi m_0^3/M m_A)~(1 - e^{-|x| m_A})$, where 
$m_A=m_0\sqrt{24\pi r/gT}$.
Our key  observation is that the gauge field mass, $m_A$, 
is inversely proportional to the
temperature. At high temperatures $m_A$ is small, and 
the potential is linear in $|x|$ upto large scales leading to a strong binding
of spinons~\cite{AO}.
However, as $T \rightarrow 0$  ~$m_A \rightarrow \infty$, and $V(x)$
reduces to a constant in which case
no bound states exist. There exists therefore a critical value of the
gauge field mass when the attraction between spinons becomes too weak to bound
them into pairs.
Careful analysis of the homogeneous version of Eq.(\ref{schrod}) performed by
Campostrini and Rossi~\cite{ital2}
 shows that bound states disappear when
$(6\pi/{M})^{1/3}~(m_0/m_A)< C$,
where $C = 0(1)$ is a numerical constant.
 Confinement-deconfinement temperature is 
then 
\begin{equation}
T^{*}= 8 C^2 (6\pi/M)^{1/3} ~\rho_{\perp}^0 r.
\end{equation}
Above $T^{*}$ there is a confinement, and the staggered static spin 
susceptibility $\chi(q)$ has a pole singularity,
$\chi_{conf}(q) = \frac{A}{q^2 + m^2}$ ($A \rightarrow 0$ as $T \rightarrow
T^*$), which translates into
$\chi (r) \propto r^{-1/2} e^{-r m}$ at large distances.
Below $T^*$, spinons are deconfined, 
and $\chi (q)$ has  only a branch-cut singularity,
 $\chi_{deconf}(q) \sim (q^2 + m^2)^{-1/2}$ which implies that at large
 distances
$\chi (r) \propto r^{-1} e^{-r m}$. There is therefore a real change in the
behavior of the physical observable at $T^*$, however, we do not expect
that there will be any changes in the thermodinamic quantities at this
temperature.

We also caution that
our low-energy analysis is
restricted to the neighborhood of the $CP^{1}$ fixed point, where the gauge
field mass is small at $T \sim T^*$. The fate of the transition line
at larger $r$ is unknown simply becase one cannot use the
Hubbard-Stratonovich decoupling when
 the gauge field mass becomes comparable to the upper cut-off of the
theory~\cite{azaria}.

The results of the present paper are in agreement with Ref~\cite{CSS}
and with the results of
other authors.
Transformation from free 
spinons at $r\neq 0$ to the confined ones at $r=0$ was observed by Wiegmann 
\cite{pasha} in the exact solution of the 2D classical $O(3)$ problem. 
Deconfinement of spinons 
in the $T=0$ quantum-disordered phase with incommensurate spin correlations,
due to the appearance of the gauge field mass, was discussed 
by Sachdev and Read \cite{sr}.
Classical $d$-dimensional versions of the actions (\ref{zzz}) and (\ref{D1})
were recently studied by Azaria et al. \cite{azaria} by the 
renormalization group and $1/M$ analysis in all dimensions between 
2 and 4.
One of their key findings is that all
models with $\gamma \neq - 1$ are asymptotically equivalent at long
distances to the $O(2M)$ model. This is in complete agreement with
our results. 
They also argued that though the low-energy physics is controlled by
$O(2M)$ fixed point, the high-energy (short distance) behavior for 
weak incommensurability is still controlled by the 
the $CP^{1}$ fixed point  - this is also consistent with our
result that confinement persists at high 
temperatures.  They, however,
did not explicitly discuss the confinement-deconfinement transition for 
the spin susceptibility.

To conclude, in this paper we analyzed the crossover from the 
$CP^{1} \equiv O(3)$ to the $O(4)$
critical behavior in a model of weakly frustrated quantum antiferromagnet.
 We have shown, in the $1/M$ expansion,
that for an arbitrary small bare longitudinal stiffness,
the system flows away from the $CP^{M-1}$ fixed 
point towards the $O(2M)$ fixed
point. The same type of critical behavior was found 
previously for the case where the bare 
longitudinal and transverse stiffnesses were close to
each other \cite{CSS}. It is therefore likely that the 
$O(2M)$ critical behavior (i.e., the $O(4)$ behavior 
for the physical case of $M=2$) holds 
 at least for all
$\rho^0_{\parallel} \leq \rho^0_{\perp}$. 
An unresolved
 issue is whether the $O(4)$ behavior holds for an arbitrary {\it large} 
ratio of $\rho^0_{\parallel}/\rho^0_{\perp}$, or there is a crossover to a
different kind of critical behavior with possible binding of spinons. This last
issue is interesting on its own grounds but it is 
also possibly related to the controversy surrounding the critical behavior of
stacked triangular antiferromagnets which, as numerical studies indicate,
possess critical exponents different from the $O(4)$ ones \cite{diep}.
 The ratio of
the stiffnesses in 2D triangular antiferromagnets  is not known exactly:
noninteracting spin-wave calculations
yield $\rho^0_{\parallel}/\rho^0_{\perp} =2$
but first $1/S$ corrections are not small and substantially reduce this ratio
 \cite{CSS2}.

We also have shown that 
 that the competition between the two zero-temperature fixed
points leads to a confinement-deconfinement transition for static spin
susceptibility at a finite temperature. 
Numerical studies of this phenomena are very desirable.

It is our pleasure to thank P. Azaria, Th. Jolicoeur, 
S. Sachdev and P. B. Wiegmann for useful 
discussions. We acknowledge support of ITP at UCSB where this work
has been completed.  A.C. is an A.P. Sloan fellow.


\begin{references}
\bibitem{comment}
Note that 
in our consideration we do not distinguish between stiffnesses and
susceptibilities, i.e., we set  $\rho^0_{\parallel} =
\chi^0_{\parallel},~\rho^0_{\perp} = \chi^0_{\perp}$. 
In general, there are two
different crossover exponents~\cite{CSS}; the term with the 
second exponent does not appear in our expression 
for ${\bar \gamma}$ 
because we set both spin-wave velocities to unity.
\end{references}
\end{document}